\documentclass[11pt, a4paper]{article}

\usepackage[utf8]{inputenc}
\usepackage[T1]{fontenc}
\usepackage{geometry}
\usepackage{amsmath, amssymb} 
\usepackage{booktabs}         
\usepackage{graphicx}         
\usepackage{longtable}        
\usepackage{enumitem}         
\usepackage{hyperref}         
\usepackage{caption}          
\usepackage{float}            

\usepackage[style=numeric, sorting=none, backend=biber]{biblatex}
\title{Synthetic Data for Veterinary EHR De-identification: Benefits, Limits, and Safety Trade-offs Under Fixed Compute}

\author{David Brundage, PhD \\
University of Wisconsin-Madison, School of Veterinary Medicine }

\date{}

\begin{document}

\maketitle

\begin{abstract}
\noindent \textbf{Background:} Veterinary electronic health records (vEHRs) are increasingly used for surveillance and research, but free-text narratives frequently contain privacy-sensitive identifiers that limit secondary use. PetEVAL provides professionally annotated veterinary clinical narratives with identifier span labels, enabling systematic evaluation of de-identification methods in this low-resource domain.

\noindent \textbf{Objective:} To evaluate when large language model (LLM)-generated synthetic veterinary narratives improve or degrade de-identification safety and utility under distinct training regimes, with particular emphasis on (i) synthetic augmentation that expands training exposure while holding real-note supervision fixed and (ii) substitution of synthetic notes for real labeled notes under a fixed training budget.

\noindent \textbf{Methods:} We conducted a controlled low-resource simulation using a PetEVAL-derived labeled corpus. The holdout evaluation set comprised 3,750 real clinical notes, and the real training subset comprised 1,249 notes. We generated a synthetic pool of 10,382 notes from masked real seeds using a constrained ``template-only'' generation regime in which identifiers were removed prior to any LLM call and placeholders were deterministically instantiated locally with exact character-offset tracking. The synthetic pool included 2,978 PII-bearing notes (containing $\ge$1 identifier span) and 7,404 no-PII notes. Three transformer backbones (PetBERT, VetBERT, Bio\_ClinicalBERT) were trained under increasing synthetic mixtures. Evaluation used seed-averaged span-level exact and overlap metrics, with document-level leakage rate—defined as the fraction of documents containing $\ge$1 missed identifier span—as the primary safety-oriented outcome.

\noindent \textbf{Results:} Under Fixed-sample substitution training, progressively substituting real clinical narratives with synthetic notes led to a monotonic increase in document-level leakage and a decline in span-level F1, indicating that synthetic data does not safely replace real supervision. Under compute-matched training, moderate synthetic mixing achieved performance comparable to real-only oversampling, while high synthetic dominance degraded utility without reducing leakage. In contrast, when training was scaled by epochs—thereby increasing total exposure—synthetic augmentation improved span-level performance and reduced document-level leakage across all backbones. For PetBERT, span-overlap F1 increased from 0.831 with real-only training to 0.850 $\pm$ 0.014 at a nominal 90\% synthetic mixture, while document-level leakage decreased from 6.32\% to 4.02\% $\pm$ 0.19\%. Similar exposure-driven trends were observed for VetBERT and Bio\_ClinicalBERT. Additional ablations showed that increasing the proportion of no-PII synthetic notes inflated precision-driven F1 but reduced recall for lower-frequency entity types, increasing document-level leakage.

\noindent \textbf{Conclusions:} LLM-generated synthetic veterinary notes can reduce document-level leakage when used to expand training exposure, but they do not substitute for labeled real notes under fixed training budgets. Compute-matched analyses indicate that observed gains largely reflect increased exposure rather than intrinsic advantages of synthetic text. Corpus diagnostics identified systematic synthetic–real mismatches—shorter note length, PER-heavy label distributions, and absence of within-note identifier repetition—that align with recall failures and persistent leakage, providing empirically grounded constraints for safer synthetic data design in veterinary de-identification.

\vspace{1em}
\end{abstract}

\section{Introduction}

\subsection{Background and Motivation}
Veterinary electronic health records (vEHRs) contain large volumes of free-text clinical narratives that are increasingly reused for surveillance, research, and quality improvement, yet remain difficult to share due to embedded owner- and clinician-identifying information. Surveillance initiatives such as the Small Animal Veterinary Surveillance Network (SAVSNET) \cite{jones_savsnet_2016} and VetCompass \cite{noauthor_vetcompass_nodate} demonstrate the population-level value of veterinary clinical text, while academic veterinary hospitals increasingly rely on narrative records to support downstream analytics for care delivery and research. However, secondary use of this data is constrained by ethical and institutional requirements to protect confidentiality. Although veterinary records are not legally covered as Protected Health Information (PHI) under HIPAA, they routinely contain identifier-like strings that must be removed to enable responsible data sharing. Manual de-identification remains the gold standard but is prohibitively expensive at scale, motivating automated natural language processing (NLP) approaches. PetEVAL addresses part of this gap by providing professionally annotated veterinary narratives with identifier span labels and a standardized evaluation protocol for veterinary de-identification models \cite{farrell-etal-2025-peteval}.

Compared with human healthcare, veterinary NLP operates in a markedly low-resource setting. Human clinical de-identification benefits from large public benchmarks such as i2b2/n2b2 \cite{stubbs_automated_2015} and extensive de-identified corpora such as MIMIC \cite{johnson_mimic-iii_2016}. In contrast, veterinary clinical text differs substantially in linguistic style—often telegraphic, species-specific, and role-dependent—requiring models to handle entities such as breeds, species, and owner–patient relationships. As a result, models pretrained on human clinical corpora frequently fail to generalize to veterinary narratives, and labeled veterinary datasets remain scarce \cite{boguslav_fine-tuning_2025}. Early veterinary NLP systems such as DeepTag \cite{nie_deeptag_2018} and VetTag \cite{zhang_vettag_2019} demonstrated the feasibility of learned approaches, but recent work has shifted toward transformer-based encoders adapted for veterinary text \cite{hur_domain_2020, farrell_petbert_2023}. These constraints motivate evaluations explicitly tailored to resource-limited medical settings \cite{posada_evaluation_2024}.

\subsection{Synthetic Data and De-identification Under Resource Constraints}
Automated de-identification has evolved from rule-based systems and dictionary matching \cite{ferrandez_evaluating_2012, stubbs_automated_2015} to statistical sequence models such as CRFs and recurrent neural networks \cite{jiang_-identification_2017, dernoncourt_-identification_2017}, and more recently to transformer-based architectures \cite{vaswani_attention_2017, johnson_deidentification_2020}. Domain-specific pretraining has consistently improved performance over general-domain adaptation, as demonstrated by BioBERT and ClinicalBERT \cite{lee_biobert_2020, alsentzer_publicly_2019, gu_domain-specific_2021}, yet comparative evaluations show that architectural advances alone do not eliminate domain mismatch \cite{meaney_comparative_2022}. In veterinary medicine, models such as VetBERT and PetBERT attempt to bridge this gap through domain-adapted pretraining \cite{hur_domain_2020, farrell_petbert_2023}, but remain constrained by limited labeled supervision.

In parallel, synthetic clinical text generation has emerged as a privacy-enhancing technology \cite{alshaikhdeeb_generation_2025}. Synthetic data is commonly used either for \emph{augmentation}, where synthetic samples expand limited training corpora, or \emph{substitution}, where synthetic data replaces real records to reduce disclosure risk \cite{li_two_2023}. Prior work in medical NLP reports that augmentation can improve named entity recognition under data scarcity \cite{chen_improved_2024, sasse_disease_2024}, but also highlights the risk that unconstrained LLM generation may memorize or hallucinate sensitive content \cite{brown_language_2020}. Template-based and masked generation strategies offer finer control over output structure and lower re-identification risk than unconstrained causal language modeling \cite{belkadi_generating_2025, zhang_training_2025}, motivating their use in safety-critical settings.

Despite increasing adoption, the utility of synthetic data for de-identification under realistic training constraints remains unclear. In particular, it is not well understood whether improvements attributed to synthetic augmentation reflect intrinsic data quality or simply increased training exposure, nor whether synthetic data can safely substitute for real labeled narratives under fixed training budgets.

\subsection{Privacy Evaluation Beyond Aggregate Metrics}
De-identification is typically evaluated as a named entity recognition task using token- or span-level precision and recall. While recall serves as a proxy for safety and precision preserves clinical utility, aggregate overlap metrics fail to capture operational risk when a single missed identifier renders an entire document unsafe for release \cite{ren_how_2025}. Broader privacy frameworks further emphasize that leakage risk persists even when average performance appears strong, motivating evaluation approaches that align more closely with release decisions \cite{ren_how_2025}.

Accordingly, we prioritize document-level leakage rate—the fraction of documents containing at least one missed identifier span—as our primary safety-oriented metric. This formulation exposes failure modes that remain hidden under aggregate scoring and is particularly relevant in low-resource settings where limited supervision amplifies the impact of recall errors.

\subsection{Gaps in Prior Work and Study Objective}
While PetEVAL provides a benchmark substrate and baseline results, it does not address how de-identification safety and utility change when labeled veterinary narratives are scarce and synthetic data are introduced. More broadly, existing evaluations rarely disentangle the effects of synthetic data composition from training regime or optimization budget, limiting interpretability \cite{heider_extensible_2024}. Evidence from human healthcare further shows that models trained on public benchmarks often fail under local documentation shifts, underscoring the need for controlled, context-aware evaluation \cite{eyre_evaluating_2025, kraljevic_validating_2023}.

In this study, we conduct a systematic evaluation of LLM-generated synthetic veterinary clinical narratives for de-identification under explicit data and compute constraints. Using a PetEVAL-derived low-resource simulation, we assess two use cases: (i) \emph{augmentation}, where synthetic data expands training exposure while real-note supervision is held fixed, and (ii) \emph{substitution}, where synthetic notes replace real labeled notes under a fixed sample budget. To minimize generation-time privacy risk, we employ a constrained template-only regime in which identifiers are removed prior to LLM prompting and instantiated deterministically with exact character-offset tracking. We evaluate three transformer backbones under exposure-scaled, fixed-sample, and compute-matched training regimes, and prioritize document-level leakage as the primary safety outcome. By explicitly linking training regime, synthetic–real distribution shifts, and document-level risk, this study provides empirically grounded constraints on when synthetic data can—and cannot—support veterinary de-identification in resource-limited settings.

\section{Methods}

\subsection{Data and Corpus Construction}
\textbf{Base corpus and study split (PetEVAL-derived low-resource simulation)} \\
This study uses a PetEVAL-derived labeled corpus. Public access to the PetEVAL corpus is currently restricted to the test partition on HuggingFace. Consequently, we could not utilize the official training/validation splits. To strictly prevent data leakage while maximizing utility, we partitioned the available test data into mutually exclusive `seed' (used for generation) and `holdout' (used for evaluation) subsets. This necessitates the `low-resource simulation' framing, as we are effectively bootstrapping a training set from a limited evaluation-only resource. The holdout evaluation set contains 3,750 real notes. The real training subset contains 1,249 notes. Because this split is created within a pre-existing PetEVAL partition, this evaluation measures in-domain augmentation utility and it should be interpreted as a controlled low-resource simulation rather than a direct replication of PetEVAL’s official cross-clinic generalisation setting. This design prioritizes internal validity of augmentation and substitution effects over external generalization.

\noindent \textbf{Corpus composition and length statistics} \\
We report note-length distributions and identifier prevalence for the real training subset, the synthetic pool, and the real holdout evaluation set (Table 1). Across subsets, the proportion of notes containing any identifiers is closely matched ($\sim$29\%), but synthetic notes are substantially shorter than real notes, and the PII-bearing synthetic notes exhibit a narrower tail in span counts per note.

\begin{table}[H]
\centering
\caption{Corpus composition and note-length statistics}
\label{tab:corpus_stats}
\resizebox{\textwidth}{!}{%
\begin{tabular}{lccccc}
\toprule
\textbf{Subset} & \textbf{N notes} & \textbf{Notes w/ $\ge$1 span} & \textbf{Words median [IQR]} & \textbf{Chars median [IQR]} & \textbf{Total spans} \\
\midrule
Train (real) & 1,249 & 362 (29.0\%) & 57 [37–83] & 333 [219–489] & 533 \\
Synthetic pool (all) & 10,382 & 2,978 (28.7\%) & 36 [27–46] & 231 [172–299] & 5,444 \\
\hspace{3mm} $\vdash$ Synthetic PII-bearing & 2,978 & 2,978 (100.0\%) & 38 [29–47] & 249 [187–305] & 5,444 \\
\hspace{3mm} $\llcorner$ Synthetic no-PII & 7,404 & 0 (0.0\%) & 35 [26–46] & 224 [167–295] & 0 \\
Holdout test (real) & 3,750 & 1,085 (28.9\%) & 57 [38–79] & 332 [220–465] & 1,569 \\
\bottomrule
\end{tabular}%
}
\end{table}

\subsection{Annotation Representation and Label Schema}
Clinical narratives are stored with raw note text in a \texttt{sentence} field and gold annotations
serialized as lists of span dictionaries with character offsets and labels
(\texttt{start}, \texttt{end}, \texttt{label}, \texttt{entity}). We operate on a compact identifier
schema comprising five labels: \texttt{PER}, \texttt{ORG}, \texttt{LOC}, \texttt{TIME}, and
\texttt{MISC}.

\paragraph{Synthetic template-to-annotation mapping.}
For synthetic data, span annotations are constructed \emph{by design} during local placeholder
instantiation rather than inferred post hoc. During generation, the large language model produces
templates containing only allowlisted placeholders (e.g., \texttt{\_\_VET1\_\_},
\texttt{\_\_CLINIC1\_\_}, \texttt{\_\_DATE1\_\_}), each of which is deterministically mapped to a
canonical identifier label via a fixed lookup table (e.g., clinician-, owner-, and patient-role
placeholders map to \texttt{PER}; clinic and laboratory placeholders map to \texttt{ORG};
city/address placeholders map to \texttt{LOC}; date/time placeholders map to \texttt{TIME};
and administrative or device identifiers map to \texttt{MISC}).

Templates are filled locally by sequentially replacing placeholders with synthetic surface forms
drawn from role-specific generators, while incrementally constructing the output string and
recording character offsets at insertion time. This procedure yields exact span boundaries and
labels without requiring heuristic span detection, XML parsing, or alignment against the rendered
text. Because placeholders are guaranteed to be unique within each template and are replaced in
a single left-to-right pass, synthetic annotations are internally consistent and free of overlap
ambiguities by construction.

\subsection{Synthetic Generation}
\textbf{Privacy-preserving templating} \\
Synthetic notes were generated using a ``template-only'' regime where real identifiers in seed notes are masked locally prior to API transmission, and the LLM is instructed to produce a JSON-formatted note template containing only allowlisted placeholders (e.g., role-aware clinician/owner/patient placeholders; clinic/organisation placeholders; date/time placeholders; location placeholders; and limited structured identifiers where applicable). This approach aligns with recent findings that template-based data generation can effectively enhance information extraction performance in specialized technical domains while maintaining structural consistency \cite{nagayama_data_2024}.
Outputs are machine-validated to enforce the allowlist and to reject any non-placeholder identifier-like content. Accepted templates are then filled locally using entity pools, while incrementally constructing the output string and recording insertion offsets to produce exact spans ``by construction.'' This constraint explicitly prevents the generation of ``hallucinated'' identifier surface forms, a failure mode recently observed when applying generative LLMs to UK clinical records \cite{kuo_benchmarking_2025}—ensuring that all synthetic PII is traceable and distinct from real patient data. To align synthetic note duration with real data, the generation prompt explicitly constrained output length to within $\pm30\%$ of the seed note length. Despite explicitly prompting for length constraints, generated notes remained consistently shorter than real narratives (median 36 vs 57 words), reflecting a known tendency of LLMs to prefer concise outputs.

\noindent \textbf{Synthetic pool structure} \\
The synthetic pool contains two subsets (Table 1):
\begin{enumerate}
    \item PII-bearing synthetic notes (n=2,978): contain $\ge$1 labeled identifier span.
    \item No-PII synthetic notes (n=7,404): contain no placeholders and are screened to avoid identifier-like strings, serving as negative examples.
\end{enumerate}
After locally instantiating placeholders (thereby producing final texts and span annotations), we applied an automated deduplication step to reduce template reuse and prevent inflated effective sample size. We removed (i) exact duplicates using a normalized text hash and (ii) near-duplicates using embedding-based cosine similarity on the final filled notes. Any candidate with cosine similarity $\ge .90$ to an existing retained note was discarded. This filtering was performed on the filled outputs (not the raw templates) so that duplicates arising from deterministic placeholder filling were also detected. We additionally filtered synthetic notes that were exact or near-duplicates of any seed note used for prompting, using the same hashing and cosine-similarity procedure, to reduce the risk of generating paraphrases that closely track source notes.

\subsection{Models and Inference Baselines}
We evaluated three transformer backbones: 
\begin{itemize}
    \item \textbf{PetBERT} (SAVSNET/PetBERT) \cite{farrell_petbert_2023}: A BERT model pretrained on a massive corpus of UK veterinary clinical narratives.
    \item \textbf{VetBERT} (havocy28/VetBERT) \cite{hur_domain_2020}: A model adapted for veterinary extracted text and clinical notes.
    \item \textbf{Bio\_ClinicalBERT} (emilyalsentzer/Bio\_ClinicalBERT) \cite{alsentzer_publicly_2019}: A domain-standard baseline pretrained on the MIMIC-III corpus.
\end{itemize}

\subsection{Training Procedure and Mixture Construction}
\textbf{Hyperparameters and Implementation Details}
To ensure reproducibility, all experiments utilized a consistent set of hyperparameters across backbones and training regimes (unless otherwise specified for compute-matched controls). Models were fine-tuned for 8 epochs with a batch size of 16 and a learning rate of 2e-5 (AdamW optimizer). We applied a weight decay of 0.01 and used mixed-precision training (FP16) to optimize throughput. To handle the class imbalance inherent in NER tasks, we applied class-weighted loss for all updates. Input sequences were truncated to a maximum length of 512 tokens, using a sliding window with an evaluation stride of 64 to handle long documents. Training included an early stopping patience of 2 epochs to prevent overfitting. While the default random seed was set to 1 for initial debugging, reported results use the multi-seed averages (3 or 5 seeds) described in the experimental design sections.\vspace{30mm}
\textbf{Experimental design A: augmentation with increasing total training size} \\
For each backbone, we swept target synthetic fractions $f \in \{0.0, 0.25, 0.5, 0.75, 0.9\}$. The training harness retains all $R$ real training notes and samples $S$ synthetic notes so that the realized synthetic fraction satisfies:
\[
\frac{S}{R+S} \approx f, \quad \text{equivalently} \quad S \approx \mathrm{round}\left(\frac{f}{1-f}R\right),
\]
capped by the available synthetic pool size. This implies that increasing $f$ increases the total number of training notes (and, under epoch-based training, the total number of parameter updates). We therefore report realized set sizes for each target fraction (Table 2).

\begin{table}[H]
\centering
\caption{Realized training sizes for the augmentation sweep (R = 1,249 real notes; synthetic pool = 10,382)}
\label{tab:realized_training}
\begin{tabular}{lcccc}
\toprule
\textbf{Target synthetic fraction} & \textbf{N train total} & \textbf{N real} & \textbf{N synthetic} & \textbf{Realized synthetic \%} \\
\midrule
0.00 & 1,249 & 1,249 & 0 & 0.0\% \\
0.25 & 1,665 & 1,249 & 416 & 25.0\% \\
0.50 & 2,498 & 1,249 & 1,249 & 50.0\% \\
0.75 & 4,996 & 1,249 & 3,747 & 75.0\% \\
0.90 & 11,631 & 1,249 & 10,382 & 89.3\% \\
\bottomrule
\end{tabular}
\end{table}

\noindent \textbf{Experimental design B: fixed-N substitution (low-resource)} \\
We held the total number of training examples constant at $N$ and varied the fraction of real labeled notes $r \in \{1.00, 0.50, 0.25, 0.10, 0.05\}$, filling the remaining $(1-r)N$ slots with synthetic examples drawn from the synthetic pool. Each condition was repeated across three random seeds. This experiment trained PetBERT with sequence length 512, eval stride 64, FP16, and class-weighted loss.

\noindent \textbf{Training-regime sensitivity and compute-matched controls} \\
To assess whether gains observed under synthetic augmentation reflect intrinsic benefits of synthetic data or increased optimization signal, we conducted two strengthening experiments. First, we repeated the epoch-based augmentation sweep across five random seeds to quantify uncertainty. Second, we introduced a compute-matched control in which all conditions were trained for a fixed number of optimizer steps, using oversampling with replacement for real-only training to match the number of examples seen. These experiments isolate the effect of synthetic data content from training-schedule confounds.

\subsection{Evaluation Protocol and Metrics}
We report:
\begin{itemize}
    \item Token-level metrics computed with seqeval.
    \item Span-level exact-match precision/recall/F1, requiring identical label and boundaries.
    \item Span-level overlap-match precision/recall/F1 using label-consistent greedy matching by maximum span intersection.
    \item Document-level leakage rate (overlap): the fraction of documents with $\ge$1 false-negative identifier span under overlap matching. We treat leakage as the primary safety-oriented metric for model selection.
\end{itemize}

\subsection{Synthetic–Real Distribution Diagnostics}
To interpret augmentation and substitution outcomes, we quantify structural differences between real and synthetic PII-bearing notes (Table 3) and overall span label distributions (Table 4). These diagnostics identify measurable distribution shifts likely to influence leakage.
The proportion of synthetic no-PII notes was chosen to mirror the identifier prevalence in the heldout evaluation set, which itself reflects the distribution of the source collection. We additionally evaluate sensitivity to this choice by varying the fraction of no-PII synthetic notes while holding total synthetic volume constant.
\begin{table}[H]
\centering
\caption{Structural differences among PII-bearing notes (percent of PII notes exhibiting each property)}
\label{tab:structural_diffs}
\resizebox{\textwidth}{!}{%
\begin{tabular}{lccccccc}
\toprule
\textbf{Subset (PII notes only)} & \textbf{PER} & \textbf{ORG} & \textbf{LOC} & \textbf{TIME} & \textbf{MISC} & \textbf{No PER} & \textbf{Any within-note entity repetition} \\
\midrule
Train (real PII notes; n=362) & 76.2 & 10.2 & 6.4 & 17.4 & 6.1 & 23.8 & 15.5 \\
Synthetic (PII-bearing; n=2,978) & 100.0 & 19.0 & 1.5 & 13.7 & 6.4 & 0.0 & 0.0 \\
Holdout test (real PII notes; n=1,085) & 76.7 & 10.1 & 5.8 & 16.9 & 6.1 & 23.3 & 14.7 \\
\bottomrule
\end{tabular}%
}
\end{table}

\begin{table}[H]
\centering
\caption{Span label distribution (counts; percent of spans in parentheses)}
\label{tab:span_labels}
\begin{tabular}{lccc}
\toprule
\textbf{Label} & \textbf{Train (real) spans} & \textbf{Synthetic pool spans} & \textbf{Holdout test spans} \\
\midrule
PER & 343 (64.4\%) & 4,229 (77.7\%) & 1,035 (66.0\%) \\
ORG & 41 (7.7\%) & 569 (10.5\%) & 121 (7.7\%) \\
LOC & 26 (4.9\%) & 46 (0.8\%) & 68 (4.3\%) \\
TIME & 98 (18.4\%) & 408 (7.5\%) & 262 (16.7\%) \\
MISC & 25 (4.7\%) & 192 (3.5\%) & 83 (5.3\%) \\
\bottomrule
\end{tabular}
\end{table}

\section{Results}

\subsection{Main Results}
\textbf{Primary ranking criterion.} We rank configurations by document-level overlap leakage (lower is better), with span-overlap F1 as a secondary utility metric. Because document-level leakage is sensitive to training regime, baseline leakage values differ across exposure-scaled, fixed-sample, and compute-matched experiments; comparisons are therefore only valid within the same training regime.

\begin{figure}[H]
    \centering
    \includegraphics[width=0.65\textwidth]{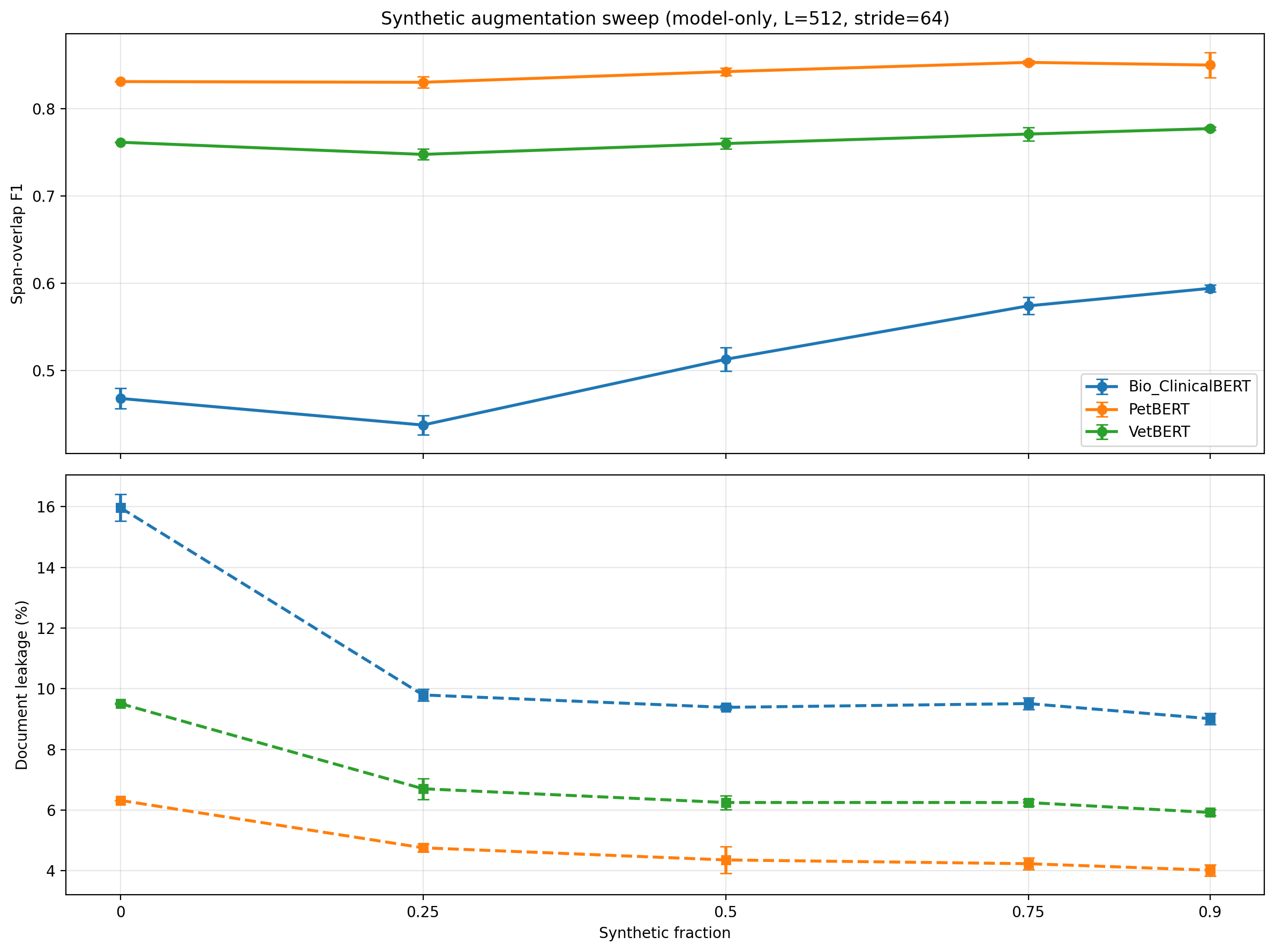}
    \caption{Synthetic augmentation sweep ($L{=}512$, stride $=64$; $n{=}3$ seeds). Points show mean; error bars show $\pm$1 SD across seeds. Top: Span-overlap F1 increases with synthetic fraction across backbones. Bottom: Document-level overlap leakage decreases with synthetic fraction, with PetBERT maintaining the lowest leakage across the sweep.}
    \label{fig:aug_sweep}
\end{figure}

\noindent \textbf{Backbone comparison at high synthetic augmentation (nominal 90\%).}
We quantified how the overall volume of synthetic augmentation impacts both utility and safety by varying the synthetic fraction from 0.0 to 0.9 while holding the training configuration constant ( $L{=}512$, stride $=64$). Across backbones, increasing synthetic fraction improved span-overlap F1 and reduced document-level overlap leakage relative to real-only training, indicating that synthetic augmentation reduces residual identifier exposure relative to real-only training under the same exposure-scaled regime.

At nominal 90\% synthetic, PetBERT achieved the best safety--utility trade-off, with mean span-overlap F1 $0.850 \pm 0.014$ and mean document-level leakage $4.02\% \pm 0.19\%$ across three seeds. VetBERT achieved mean F1 $0.777 \pm 0.002$ with leakage $5.92\% \pm 0.10\%$, while Bio\_ClinicalBERT lagged with mean F1 $0.594 \pm 0.004$ and leakage $9.01\% \pm 0.18\%$ (Table~\ref{tab:best_backbone}). Seed-to-seed variability was small compared to between-backbone differences, supporting that the backbone ranking is not initialization-specific.

Consistent with a risk-oriented interpretation, improvements in F1 do not fully characterize safety: leakage can plateau at high synthetic fractions even as F1 continues to increase, suggesting diminishing safety returns once frequent identifier patterns are sufficiently covered. In contrast, the substantially higher leakage of Bio\_ClinicalBERT across mixture levels indicates that augmentation alone does not fully compensate for domain mismatch in safety-critical behavior.

Across all backbones, document-level leakage decreased under Exposure-scaled (epoch-based) synthetic augmentation but increased monotonically under fixed-sample, indicating that synthetic data improves coverage only when it expands training exposure rather than replacing real supervision.”

\begin{table}[H]
\centering
\caption{Backbone comparison at nominal 90\% synthetic ( $L{=}512$, stride $=64$; $n{=}3$ seeds). Ranked by mean document-level overlap leakage (lower is better).}
\label{tab:best_backbone}
\resizebox{\textwidth}{!}{%
\begin{tabular}{lcccc}
\toprule
\textbf{Rank} & \textbf{Backbone} & \textbf{Target synthetic fraction} &
\textbf{Span-overlap F1 (mean $\pm$ SD)} & \textbf{Doc leakage (mean $\pm$ SD)} \\
\midrule
1 & PetBERT & 0.90 & $0.850 \pm 0.014$ & $4.02\% \pm 0.19\%$ \\
2 & VetBERT & 0.90 & $0.777 \pm 0.002$ & $5.92\% \pm 0.10\%$ \\
3 & Bio\_ClinicalBERT & 0.90 & $0.594 \pm 0.004$ & $9.01\% \pm 0.18\%$ \\
\bottomrule
\end{tabular}%
}
\vspace{0.5em}
\footnotesize{
Leakage is the fraction of documents with at least one missed identifier span under overlap matching. Results correspond to Exposure-scaled (epoch-based) training with three random seeds; real-only baselines in this table are not directly comparable to Fixed-sample substitution (fixed-N) or compute-matched experiments reported elsewhere.
}
\end{table}

\begin{figure}[H]
    \centering
    \includegraphics[width=0.9\textwidth]{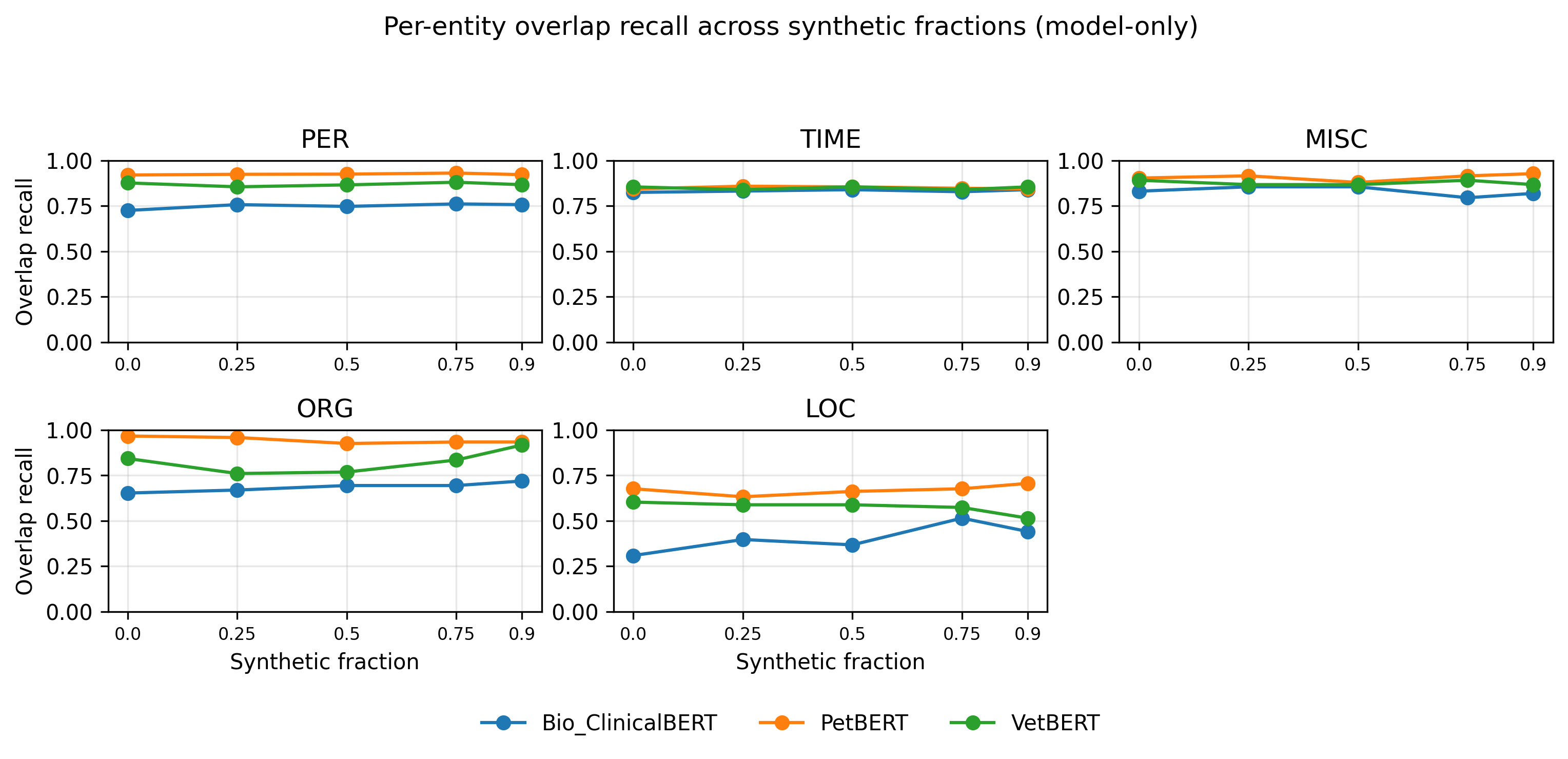}
    \caption{Per-entity overlap recall across synthetic fractions. Synthetic augmentation drives recall gains in minority classes (e.g., LOC/ORG) while high-frequency classes (PER) change modestly.}
    \label{fig:entity_recall}
\end{figure}

\noindent \textbf{Synthetic augmentation improves PetBERT performance.}
For PetBERT, increasing the synthetic fraction from 0.0 to nominal 0.9 increased mean span-overlap F1 from $0.831$ (real-only) to $0.850 \pm 0.014$, while mean document-level leakage decreased from $6.32\%$ to $4.02\% \pm 0.19\%$ across seeds. Improvements were most visible in recall for lower-frequency entity types (Figure~\ref{fig:entity_recall}), consistent with the observed reductions in document-level leakage. We note that the real-only PetBERT baseline leakage differs across experimental settings. Under the primary Exposure-scaled (epoch-based) training protocol used for the main augmentation results (Table 5, Figure 1), the real-only baseline exhibited a document-level overlap leakage of 6.32. In contrast, under alternative robustness and compute-matched protocols that fix the number of optimization steps or total training examples (Tables 6–8), the real-only baseline leakage was lower, ranging from 3.50\% to 3.81\%. These differences reflect changes in training regime rather than data composition and reduce the absolute magnitude of safety improvements attributable to synthetic augmentation under fixed-compute or Fixed-sample substitution (fixed-N) conditions.

\noindent \textbf{Fixed-N substitution: synthetic fill does not fully replace real labels.} With total training size held constant, span-overlap F1 declined modestly as real fraction decreased: 0.847 (100\% real), 0.843 (50\% real), 0.830 (25\% real), 0.817 (10\% real), 0.820 (5\% real). Critically, document-level overlap leakage rose monotonically from 3.64\% (100\% real) to 5.45\% (5

\begin{table}[H]
\centering
\caption{Fixed-N substitution results (PetBERT; means over 3 seeds)}
\label{tab:fixed_n}
\resizebox{\textwidth}{!}{%
\begin{tabular}{lccc}
\toprule
\textbf{\% Real in fixed-N train} & \textbf{Span overlap F1 (mean)}  & \textbf{Doc leakage (overlap)} \\
\midrule
100\% & 0.847 & 3.64\% \\
50\% & 0.843 & 4.16\% \\
25\% & 0.830 & 4.69\% \\
10\% & 0.817 & 5.34\% \\
5\% & 0.820 & 5.45\% \\
\bottomrule
\end{tabular}%
}
\vspace{0.5em}
\footnotesize{These experiments use Fixed-sample substitution or compute-matched training regimes; real-only baseline leakage values are therefore not directly comparable to Exposure-scaled (epoch-based) results in Table 5.
}
\end{table}

\noindent \textbf{Error mode driving leakage.} Across fixed-N conditions, degradation was primarily recall-driven: span-overlap recall decreases as real supervision shrinks, while span-overlap precision remains comparatively stable. This pattern is consistent with the leakage increase, since leakage is triggered by any false-negative span within a document.

\subsection{Error Analysis}
Qualitative inspection identified three recurring phenomena relevant to de-identification safety:
\begin{itemize}
    \item \textbf{Boundary sensitivity:} punctuation and irregular formatting often yield overlap hits but exact-match failures. Because redaction is more tolerant to minor boundary mismatches than missed spans, reporting both exact and overlap criteria is informative.
    \item \textbf{Semantic ambiguity and label confusions:} some terms can reasonably be interpreted as person-like, organisation-like, or location-like depending on context. In a safety-oriented de-identification setting, span detection (even with label ambiguity) is less critical than omission.
    \item \textbf{Gold incompleteness and over-redaction:} models sometimes flag plausible identifier-like spans absent from gold annotations. Over-redaction can be acceptable for privacy protection but complicates metric-only comparisons; targeted adjudication on sampled disagreements is necessary to quantify annotation noise and its effect on reported precision.
\end{itemize}

\subsection{Training-Regime Sensitivity of Synthetic Augmentation}
To evaluate whether performance improvements attributed to synthetic augmentation are driven by synthetic data content or by increased optimization signal, we compared standard epoch-based training with compute-matched training that fixes the number of optimizer updates.
\begin{figure}[H]
    \centering
    \includegraphics[width=\textwidth]{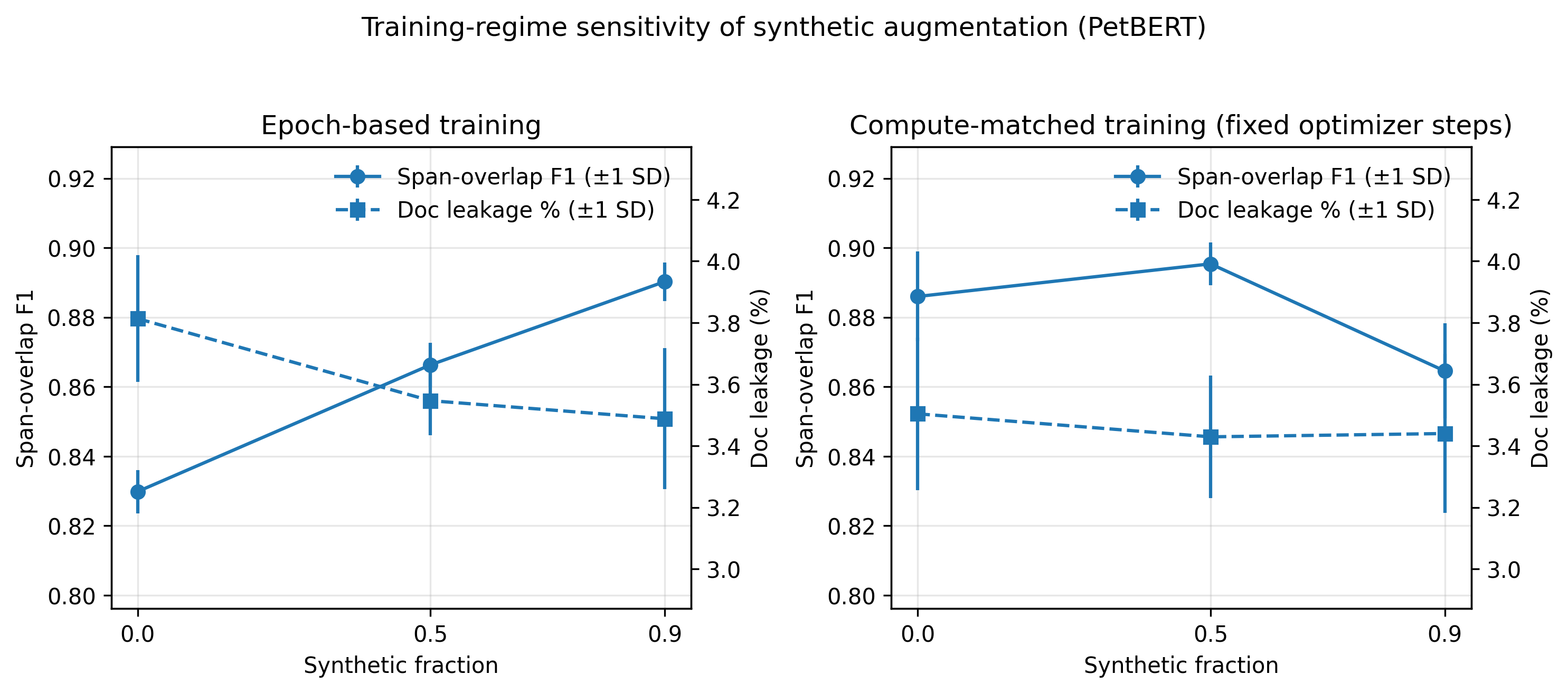}
    \caption{Training-regime sensitivity of synthetic augmentation (PetBERT). Left: Epoch-based training shows monotonic improvement in F1 and reduction in leakage. Right: Compute-matched training (fixed optimizer steps) shows performance peaks at moderate synthetic mixing (50\%) rather than 90\%.}
    \label{fig:training_regime}
\end{figure}
\subsubsection{Epoch-based augmentation with uncertainty}
Under epoch-based training, increasing the synthetic fraction consistently improved both utility and safety metrics for PetBERT across five random seeds. Mean span-overlap F1 increased from 0.830 at 0\% synthetic to 0.866 at 50\% synthetic and 0.890 at 90\% synthetic, while document-level overlap leakage decreased from 3.81\% to 3.49\%. Improvements were most pronounced for minority entity types, particularly LOC, consistent with the main augmentation results reported in \S3.1.(Table~\ref{tab:e2_epoch_uncertainty})

\begin{table}[H]
\centering
\caption{Epoch-based augmentation with uncertainty (PetBERT, 5 seeds).}
\label{tab:e2_epoch_uncertainty}
\resizebox{\textwidth}{!}{%
\begin{tabular}{lcccc}
\toprule
\textbf{Synthetic fraction} &
\textbf{Span-overlap F1 (mean $\pm$ SD)} &
\textbf{95\% CI (F1)} &
\textbf{Doc leakage (\% mean $\pm$ SD)} &
\textbf{95\% CI (leakage)} \\
\midrule
0.0 & $0.830 \pm 0.006$ & $\pm 0.0055$ & $3.81 \pm 0.21$ & $\pm 0.18$ \\
0.5 & $0.866 \pm 0.006$ & $\pm 0.0055$ & $3.55 \pm 0.11$ & $\pm 0.10$ \\
0.9 & $0.890 \pm 0.006$ & $\pm 0.0049$ & $3.49 \pm 0.23$ & $\pm 0.20$ \\
\bottomrule
\end{tabular}%
}
\vspace{0.5em}
\footnotesize{
Values are reported as mean $\pm$ standard deviation across five random seeds.
Confidence intervals are 95\% normal-approximation intervals ($\pm 1.96 \cdot \mathrm{SD} / \sqrt{n}$).
Document-level leakage is the fraction of documents with at least one missed identifier span under overlap matching. These experiments use Fixed-sample substitution or compute-matched training regimes; real-only baseline leakage values are therefore not directly comparable to Exposure-scaled (epoch-based) results in Table 5.
}
\end{table}

\subsubsection{Compute-matched (fixed-steps) augmentation}
When training was constrained to a fixed number of optimizer steps, the effect of synthetic augmentation changed qualitatively. Under this compute-matched regime, moderate synthetic mixing (50\%) achieved the best overall performance, while very high synthetic dominance (90\%) degraded span-overlap F1 relative to real-only oversampled training. Document-level leakage remained similar across compute-matched conditions (3.43–3.50\%), indicating that excessive synthetic dominance primarily affected generalization and precision rather than safety.
Per-label analysis revealed modest improvements in LOC recall at moderate synthetic fractions, but reductions in TIME and PER precision at high synthetic fractions, consistent with the synthetic–real distribution shifts described in \S2.7.(Table~\ref{tab:e1_compute_matched})

\begin{table}[H]
\centering
\caption{Compute-matched (fixed-steps) augmentation (4{,}000 optimizer steps, PetBERT, 5 seeds).}
\label{tab:e1_compute_matched}
\resizebox{\textwidth}{!}{%
\begin{tabular}{lcccccc}
\toprule
\textbf{Synthetic fraction} &
\textbf{Span-overlap F1 (mean $\pm$ SD)} &
\textbf{95\% CI (F1)} &
\textbf{Precision (mean $\pm$ SD)} &
\textbf{Recall (mean $\pm$ SD)} &
\textbf{Doc leakage (\% mean $\pm$ SD)} &
\textbf{95\% CI (leakage)} \\
\midrule
0.0 (real-only) &
$0.886 \pm 0.013$ & $\pm 0.011$ &
$0.870 \pm 0.028$ & $0.903 \pm 0.007$ &
$3.50 \pm 0.25$ & $\pm 0.22$ \\

0.5 &
$\mathbf{0.895 \pm 0.006}$ & $\pm 0.005$ &
$0.886 \pm 0.010$ & $0.905 \pm 0.004$ &
$\mathbf{3.43 \pm 0.20}$ & $\pm 0.17$ \\

0.9 &
$0.865 \pm 0.014$ & $\pm 0.012$ &
$0.828 \pm 0.019$ & $0.904 \pm 0.009$ &
$3.44 \pm 0.26$ & $\pm 0.23$ \\
\bottomrule
\end{tabular}%
}
\vspace{0.5em}
\footnotesize{
All conditions were trained for an identical number of optimizer steps.
The real-only baseline was oversampled with replacement to match the number of examples seen.
Bold indicates the best mean performance under compute-matched training. These experiments use Fixed-sample substitution or compute-matched training regimes; real-only baseline leakage values are therefore not directly comparable to Exposure-scaled (epoch-based) results in Table 5.
}
\end{table}

\subsection{Effect of no-PII synthetic proportion on de-identification safety.}
We evaluated the impact of varying the proportion of no-PII synthetic notes while holding the total synthetic volume fixed (synthetic fraction = 0.75) using PetBERT and a fixed training configuration. Increasing the fraction of no-PII synthetic examples led to a monotonic increase in span-overlap F1, peaking at 0.888 when all synthetic notes contained no identifiers, reflecting improved precision. However, this gain did not translate to improved safety. Document-level overlap leakage—our primary risk-oriented metric—was lowest when synthetic data consisted entirely of PII-bearing notes (3.09\%) and increased as the no-PII proportion grew, reaching 3.47\% at 75\% no-PII and 3.39\% at 100\% no-PII.

This degradation was recall-driven: span-overlap recall peaked at a moderate no-PII proportion (50\%) and declined thereafter, particularly for lower-frequency entity types such as TIME and LOC, which disproportionately contribute to document-level leakage. These results demonstrate a clear divergence between aggregate F1 and safety-critical behavior: excessive no-PII augmentation improves apparent utility while increasing the probability of leaking at least one identifier per document. Based on this trade-off, we select a balanced mixture (50\% no-PII synthetic) for subsequent experiments, as it achieves near-minimal leakage while preserving recall.

To assess robustness, we repeated this experiment across five random seeds at the selected operating point (50\% no-PII synthetic, synthetic fraction = 0.75). Across runs, document-level overlap leakage remained stable (mean 3.5\% ± 0.1\%), and span-overlap recall and F1 exhibited low variance, indicating that the observed trade-off between no-PII proportion and de-identification safety is not an artifact of a particular initialization. These results support the use of a balanced no-PII mixture as a stable operating regime rather than a single-seed optimum.

\subsubsection{Summary}
Collectively, these experiments demonstrate that the benefits of synthetic augmentation are constrained by both training regime and data composition. While augmentation drives monotonic improvements under epoch-based scaling by increasing effective training signal, fixed-compute controls reveal that these benefits peak at moderate mixing ratios. Furthermore, data composition directly impacts the safety-utility trade-off: increasing the proportion of 'no-PII' negative examples improves aggregate precision but degrades safety-critical recall, necessitating a balanced mixture to minimize document-level leakage.

\section{Discussion}

\subsection{Principal Findings}
This study evaluated the role of LLM-generated synthetic data in veterinary clinical de-identification under a controlled low-resource simulation, reflecting the operational constraints faced by many veterinary institutions. Unlike human healthcare, which benefits from large publicly available de-identification benchmarks and corpora, veterinary informatics lacks extensive labeled resources, making annotation cost a primary barrier to deployment. Within this context, our results clarify the conditions under which synthetic data can support de-identification model development—and, critically, the conditions under which it fails.

Across experiments, the primary benefit of synthetic augmentation was not improved optimization efficiency per se, but increased exposure to identifier patterns that are underrepresented in small real training sets. When synthetic data was used to expand the effective training corpus under epoch-based scaling, span-level performance improved and document-level leakage decreased across transformer backbones. However, this improvement was not intrinsic to synthetic text. Compute-matched controls showed that most gains attributed to high synthetic fractions attenuated or disappeared when the number of optimizer updates was held constant, and that excessive synthetic dominance could degrade utility without reducing leakage. These findings indicate that exposure-driven effects, rather than synthetic fidelity alone, account for much of the observed improvement.

Importantly, synthetic data did not safely substitute for real labeled supervision. Under Fixed-sample substitution training, progressively replacing real notes with synthetic examples produced a monotonic increase in document-level leakage, even when aggregate F1 remained comparatively high. This divergence underscores a key limitation of conventional span-level metrics for safety-critical de-identification tasks: models may achieve strong average performance while still leaving a non-trivial fraction of documents unsafe for release. Our results align with prior evidence in clinical NLP showing that aggregate metrics can obscure privacy-relevant failure modes and reinforce the necessity of document-level evaluation for operational decision-making.

Together, these findings support a conditional interpretation of synthetic augmentation in veterinary de-identification. Synthetic data can be beneficial when it expands training exposure and selectively improves recall for rare identifier patterns, but it does not replace real supervision and offers limited advantage under fixed compute constraints. Synthetic augmentation should therefore be viewed as complementary rather than substitutive, and its evaluation must account explicitly for training regime and optimization budget.

\subsection{Synthetic--Real Distribution Shifts as a Mechanistic Explanation}
Corpus diagnostics revealed systematic distribution shifts between real and synthetic notes that plausibly explain the observed substitution gap. First, a pronounced \textbf{length shift} was evident: synthetic notes were substantially shorter than real narratives (median 231 vs.\ 333 characters). If longer notes concentrate residual identifiers or present more complex boundary conditions, this shift reduces exposure to long-range contexts that contribute disproportionately to document-level leakage. Notably, this length discrepancy persisted despite explicit length-control prompting, suggesting inherent limitations in unconstrained LLM verbosity.

Second, a \textbf{label mix shift} was observed, with synthetic spans dominated by person identifiers and comparatively sparse coverage of time and location entities. A PER-heavy synthetic pool can inflate apparent performance on common identifiers while leaving lower-frequency but leakage-critical patterns undertrained. Third, a \textbf{structural shift} was present within PII-bearing notes: while all synthetic PII-bearing notes contained at least one PER span, approximately one quarter of real PII-bearing notes contained no PER at all, indicating underrepresentation of ``date-only'' or ``organisation-only'' identifier configurations. Finally, synthetic notes exhibited a complete absence of \textbf{within-note identifier repetition}, whereas repeated mentions occurred in a meaningful fraction of real notes, reducing training signal for recovery of repeated identifiers.

These measurable shifts provide a mechanistic explanation for why synthetic substitution fails and why gains under augmentation plateau. They also suggest concrete directions for improving synthetic data utility, including targeted prompting for PER-absent notes, longer narrative contexts, enhanced LOC/TIME coverage, and explicit modeling of repeated-mention patterns.

\subsection{Limitations}
This study was conducted as an in-domain low-resource simulation using a PetEVAL-derived split rather than the official cross-clinic generalization setting. While this design reflects the cold-start conditions faced by many veterinary institutions, the results should be interpreted as evidence of in-domain augmentation behavior rather than cross-site generalizability. Additionally, the synthetic generation regime was intentionally conservative to avoid direct exposure of real identifiers during generation, which may bias synthetic outputs toward cleaner and more regular structures than naturally occurring vEHR narratives.

Our analysis focused on encoder-based sequence labeling models rather than generative de-identification systems. This choice reflects the safety-critical requirement that non-sensitive clinical text remain unmodified during redaction, but it does not address potential hybrid or post-editing approaches. Finally, while compute-matched controls mitigate major confounds, future work should explore a broader range of optimization budgets and architectural variants, including long-context encoders, to further characterize exposure–utility trade-offs.

\subsection{Implications for Veterinary De-identification}
For deployment-oriented veterinary de-identification, these findings indicate that synthetic augmentation must be paired with a minimum quantity of real labeled data and evaluated using safety-oriented metrics that reflect document-level risk. Naïve substitution of real notes with synthetic data is likely to increase leakage, even when span-level performance appears strong. More broadly, our results demonstrate that assessments of synthetic data utility must be compute-aware to avoid conflating increased training exposure with intrinsic data quality.

By explicitly linking training regime, data composition, and document-level leakage, this study provides empirically grounded constraints for the responsible use of synthetic data in veterinary de-identification and highlights the need for evaluation frameworks that prioritize safety over aggregate performance metrics.

\section{Conclusion}
LLM-generated synthetic veterinary clinical notes, produced via privacy-preserving template generation and deterministic local placeholder filling, can reduce document-level leakage when they expand training exposure under epoch-based training. However, under a fixed training budget, replacing real labeled notes with synthetic notes increases document-level leakage, indicating that synthetic data is complementary rather than substitutive for real annotation in safety-critical veterinary de-identification. Importantly, we show that the benefits of synthetic augmentation are sensitive to training regime and compute constraints, suggesting the need for compute-aware evaluation when assessing synthetic data utility. Quantitative corpus diagnostics show that synthetic notes differ from real notes in length, label mix, and within-note repetition, identifying key distribution shifts that must be addressed for next-generation synthetic data design focused on reducing leakage rather than maximizing aggregate F1.

\section*{Artificial Intelligence Disclosure (AID) Statement}

\textbf{Artificial Intelligence Tool(s):} ChatGPT and Gemini 3 Pro (used between December 2025 and January 2026);
\textbf{Methodology and Execution:} AI tools were used to support controlled synthetic generation of veterinary clinical narratives as part of the de-identification experiments, with all study design, annotation schemas, and evaluation procedures defined and validated by the authors;
\textbf{Data Curation and Privacy:} No real identifiable data were shared with AI systems; all synthetic outputs were programmatically filtered and structured, and all real veterinary EHR data were handled in accordance with institutional data governance and ethical guidelines;
\textbf{Writing—Review \& Editing:} AI tools were used for limited sentence-level editing during manuscript revision, with all content reviewed and approved by the authors.

\printbibliography

\end{document}